\documentclass{PoS}
\usepackage{amsmath}
\usepackage{graphicx}
\usepackage{psfrag}
\usepackage[sort&compress,numbers]{natbib}

\def\qb{{\overline{q}}}
\def\db{\overline{d}}
\def\ub{\overline{u}}
\def\sb{\overline{s}}

\def\eps{\epsilon}
\def\NGluon{{\tt NGluon}~}

\title{One-Loop Amplitudes for Multi-Jet Production at Hadron Colliders}

\ShortTitle{Multi-Jet Amplitudes}

\author{\speaker{Simon Badger}\\
The Niels Bohr International Academy and Discovery Center, %
The Niels Bohr Institute, %
University of Copenhagen, %
Blegdamsvej 17, DK-2100 Copenhagen \O, Denmark\\
        E-mail: \email{badger@nbi.dk}}

\author{Benedikt Biedermann\\
Humboldt-Universit\"at zu Berlin, Institut f\"ur Physik, %
Newtonstra{\ss}e 15, D-12489 Berlin, Germany\\
        E-mail: \email{benedikt.biedermann@physik.hu-berlin.de}}

\author{Peter Uwer\\
Humboldt-Universit\"at zu Berlin, Institut f\"ur Physik, %
Newtonstra{\ss}e 15, D-12489 Berlin, Germany\\
        E-mail: \email{peter.uwer@physik.hu-berlin.de}}

\abstract{We present a numerical implementation for virtual corrections to
multi-jet production at Next-to-Leading order. Using the algorithm of
generalised unitarity we compute primitive amplitudes from tree-level input.
These basic ingredients are then used to compute full colour and helicity
summed corrections.}

\FullConference{ 10th International Symposium on Radiative Corrections
(Applications of Quantum Field Theory to Phenomenology) - Radcor2011\\
September 26-30, 2011\\ Mamallapuram, India}

\begin{document}

\section{Introduction}

Next-to-leading order (NLO) corrections to observables with multi-particle
final states allow for precise predictions of the complicated signal and background
reactions measured at the Large Hadron Collider (LHC) at CERN. Recent years
have seen rapid progress in the development of theoretical tools to handle the
computation, and automation, of these challenging processes. On-shell approaches have
been particularly successful at pushing the limit of multiplicity with predictions
for $2\to5$ or higher processes being achieved \cite{Frederix:2010ne,Berger:2010zx,Ita:2011wn,Becker:2011vg}.
Multi-jet corrections at NLO present an additional level of difficulty in the larger number
of parton level processes in both virtual and real radiation contributions, nevertheless full
NLO predictions have recently been achieved for a four jet final state \cite{Bern:2011ep}.

New developments are continuing to improve the algorithms and open up the range
and flexibility of available predictions with a large number under discussion at
this conference \cite{Becker:2011aa,Cascioli:2011va,Campanario:2011ud,Campanario:2011cs,Denner:2010jp,Bevilacqua:2010qb,Bevilacqua:2011xh,
Mastrolia:2010nb,Cullen:2011ac,Hirschi:2011pa,Hirschi:2011rb,Melia:2011dw,Melia:2011gk,Melia:2010bm}

In this conference note we present some new developments in the computation of
one-loop multi-parton processes using the \NGluon \cite{Badger:2010nx} c++ framework. The numerical
library uses on-shell methods to compute the basic building blocks of the
virtual contributions at NLO which we summarise in section
\ref{sec:primitives}. We then look at the construction of full colour and
helicity summed interference with the born level amplitudes in section
\ref{sec:colour} before presenting some results of some performance tests in
section \ref{sec:performance}. Finally we outline some future directions in the
conclusions.

\section{Primitive Amplitudes with \NGluon \label{sec:primitives}}

The basic building blocks, the primitive amplitudes, of our one-loop amplitudes
are computed using the generalised unitarity algorithm %
\cite{Bern:1994zx,Bern:1994cg,Ossola:2006us,Forde:2007mi,Ellis:2007br,Giele:2008ve,%
Anastasiou:2006jv,Anastasiou:2006gt,Badger:2008cm} (for recent reviews on the subject see e.g. \cite{Britto:2010xq,Ellis:2011cr,Ita:2011hi}) implemented into \NGluon.
The procedure extracts the coefficients of the scalar integral basis from
products of one-shell tree-level amplitudes which are computed using
Berends-Giele recursion relations. Some new features including multiple fermion
primitive amplitudes and speed improvements through caching and re-using common
objects in helicity and permutation sums have been described in a recent
conference note \cite{Badger:2011zv}. The scalar loop integrals functions have
been implemented in a number of public codes
\cite{vanOldenborgh:1990yc,vanHameren:2010cp,Ellis:2007qk}, the default choice
in \NGluon is {\tt FF/QCDLoop}. The procedure follows in a similar vein to other successful approaches
treating multi-gluon amplitudes \cite{Berger:2008sj,Giele:2008bc,Lazopoulos:2008ex,Giele:2009ui,Giele:2011tm}.

\begin{figure}[b!]
  \begin{center}
    \psfrag{1}{$1$}
    \psfrag{2}{$2$}
    \psfrag{3}{$3$}
    \psfrag{4}{$4$}
    \psfrag{l0}{$l_0$}
    \psfrag{P1}{$A^{[m]}_n(1_d,2_{\ub},3_u,4_{\db},\dots)$}
    \psfrag{P2}{$A^{[f]}_n(1_u,2_{\ub},3,4,\dots)$}
    \includegraphics[width=\textwidth]{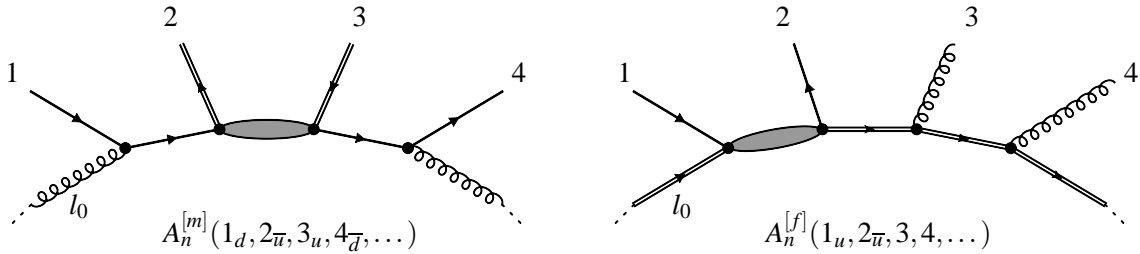}
  \end{center}
  \caption{Examples of primitive amplitudes}
  \label{fig:prim}
\end{figure}

An $n$-particle primitive amplitude is defined by an ordered set of $n$
external particles and $n$ internal propagators which can be represented by a
parent diagram. The full set of independent primitives fall into two
categories:
\begin{itemize}
  \item Amplitudes with a mixture of quark and gluon propagators in the loop, $A^{[m]}_n$.
  \item Amplitudes with an internal fermion loop, $A^{[f]}_n$.
\end{itemize}
Each parent diagram can be uniquely specified by a particle in the first
position and the propagator immediately before hand, a gluon for the mixed case
and a fermion for the fermion loop. Each subsequent propagator will be
determined by the external particle and propagator appearing before it, either
a gluon, quark or blank/dummy propagator in the case that no vertex exists.
Some examples are shown in figure \ref{fig:prim}.

The on-shell unitarity method employed in \NGluon uses the Four-Dimensional
Helicity (FDH) scheme. All tree level amplitudes can be computed in four dimensions
using the spinor helicity formalism with rational terms extracted from mass shifted 
fermion and scalar amplitudes.

A strong cross check on the validity of the rational terms comes from
non-trivial relations which lead to cancellations in super-symmetric Yang-Mills
theories \cite{Bern:1994zx,Bern:1994cg,Bern:1994fz}. For the pure gluonic amplitudes, or those with a single fermion pair,
the rational parts, $\mathcal{R}$, obey:
\begin{align}
  &\mathcal{R}\left( A_n^{[m]}(1,\ldots,n) - A_n^{[f]}(1,\ldots,n) \right) = 0, \\
  &\mathcal{R}\Big( 
  A_n^{[m]}(1_\qb,2,\ldots,(k-1),k_q,(k+1)\ldots,n) 
  + (-1)^n A_n^{[m]}(1_\qb,n,\ldots,(k+1),k_q,(k-1),\ldots,2) \nonumber\\&
  - A_n^{[f]}(1_\qb,2,\ldots,(k-1),k_q,(k+1)\ldots,n)
  -(-1)^n A_n^{[f]}(1_\qb,n,\ldots,(k+1),k_q,(k-1),\ldots,2)
  \Big) = 0.
  \label{eq:SUSY}
\end{align}
Since the rational terms are the most expensive parts of the computation of the
primitive amplitudes using these relations to reduce the total number of
rational terms leads to a considerable improvement in evaluation time.

\section{Colour Summed Amplitudes \label{sec:colour}}

The full colour amplitudes can be written as a decomposition of colour factors
and partial amplitudes,
\begin{equation}
  \mathcal{A}^{(l)}(\{p\},\{h\}) = \sum_k C_k(\{a\},\{i\}) A^{(l)}_k(\{p\},\{h\}),
  \label{eq:coldecomp1}
\end{equation}
where $l$ is the loop order.

In the above $\{i\}$ and $\{a\}$ represent fundamental and adjoint $SU(N_c)$ indices
respectively while $\{p\}$ and $\{h\}$ represent the momenta and helicity.  The one-loop
partial amplitudes must be further decomposed into the primitive objects
described in the previous section.  For the pure gluonic case and the case of a
single fermion pair such a decomposition is known to all-orders. As an explicit
example, the gluonic amplitudes are written as,
\begin{align}
  \mathcal{A}_n^{(0)} &= \sum_{\sigma \in S_{n-1}/\mathbb{Z}} tr(1,\sigma_2,\ldots,\sigma_n) A^{(0)}_n(1,\sigma_2,\ldots,\sigma_n)
  \\
  \mathcal{A}_n^{(1)} &= 
  \sum_{\sigma \in S_{n-1}/\mathbb{Z}} N_c tr(1,\sigma_2,\ldots,\sigma_n) A^{(1)}_{n;1}(1,\sigma_2,\ldots,\sigma_n) \nonumber\\
  &+\sum_{c=3}^{[n/2]-1} \sum_{\sigma \in S_{n}/S_{n;c}} 
  tr(1,\sigma_2,\ldots,\sigma_{c-1}) tr(\sigma_c,\ldots,\sigma_n) A^{(1)}_{n;c}(1,\sigma_2,\ldots,\sigma_n),
\end{align}
where
\begin{align}
  A^{(1)}_{n;1}(1,\ldots,n) &= A^{[m]}_n(1,\ldots,n) - \frac{N_f}{N_c} A^{[f]}_n(1,\ldots,n), \\
  A^{(1)}_{n;c}(1,\{\alpha\},\{\beta\}) &= (-1)^n\sum_{\sigma \in {\rm COP}\{\alpha\}^T\{\beta\}} A^{(1)}_{n;1}(1,\sigma_2,\ldots,\sigma_n)
  \label{eq:gluonpartials}
\end{align}
The sum `$\rm COP$' is over all possible mergings of the sets $\{\alpha\}=\{2,\ldots,c-1\}$ and $\{\beta\}=\{c,\ldots,n\}$ while keeping $\alpha^T$
fixed and using the cyclically ordered permutations of $\beta$.  For the
multi-fermion amplitudes the decompositions can be constructed by a systematic
matching of terms to a Feynman diagram representation \cite{Ellis:2011cr,Ita:2011ar}.

The NLO corrections come from the interference between tree-level and
loop-level amplitudes,
\begin{equation}
  d\sigma^{\rm virtual} = 2 Re(\mathcal{A}^{(0),\dagger}\cdot\mathcal{A}^{(1)}) 
  \label{eq:NLOvirt}
\end{equation}
Since the number of colour structures becomes extremely large with the
increasing multiplicity it is important to reduce the colour basis as much as
possible. We apply photon decoupling identities (and their generalisations to
the multi-fermion case) in order to reduce the amplitudes to an $(n-2)!$ of
tree amplitudes. As noted by Dixon, Del Duca and Maltoni \cite{DelDuca:1999rs}, this has some rather
striking simplification in the final representation of eq. (\ref{eq:NLOvirt}).

\begin{table}[h]
\centering
  \begin{tabular}[h]{|l|c|c|c|}
  \hline
  & $A^{[0]}_n$ & $A^{[m]}_n$ & $A^{[f]}_n$ \\
  \hline
  $\mathcal{A}_4(g,g,g,g)$ & 2 & 3 & 3 \\
  \hline
  $\mathcal{A}_4(\db,d,g,g)$ & 2 & 6 & 0 \\
  \hline
  $\mathcal{A}_4(\db,d,\ub,u)$ & 1 & 4 & 1 \\
  \hline
\end{tabular}
\begin{tabular}[h]{|l|c|c|c|}
  \hline
  & $A^{[0]}_n$ & $A^{[m]}_n$ & $A^{[f]}_n$ \\
  \hline
  $\mathcal{A}_5(g,g,g,g,g)$ & 6 & 12 & 12 \\
  \hline
  $\mathcal{A}_5(\db,d,g,g,g)$ & 6 & 24 & 9 \\
  \hline
  $\mathcal{A}_5(\db,d,\ub,u,g)$ & 3 & 16 & 3 \\
  \hline
\end{tabular}\\
\begin{tabular}[h]{|l|c|c|c|}
  \hline
  & $A^{[0]}_n$ & $A^{[m]}_n$ & $A^{[f]}_n$ \\
  \hline
  $\mathcal{A}_6(g,g,g,g,g,g)$ & 24 & 60 & 60 \\
  \hline
  $\mathcal{A}_6(\db,d,g,g,g,g)$ & 24 & 120 & 59 \\
  \hline
  $\mathcal{A}_6(\db,d,\ub,u,g,g)$ & 12 & 80 & 14 \\
  \hline
  $\mathcal{A}_6(\db,d,\ub,u,\sb,s)$ & 4 & 32 & 4 \\
  \hline
\end{tabular}

\caption{The number of independent primitive amplitudes appearing in the colour sums at tree-level ($A^{[0]}_n$) 
and at one-loop for the mixed ($A^{[m]}_n$) and fermion loop ($A^{[f]}_n$) cases. Like-flavour amplitudes for multiple fermions
are obtained by (anti-)symmetrization and therefore contain larger bases of primitives.}
\label{tab:primtives}
\end{table}

\section{Performance Tests \label{sec:performance}}

\subsection{Infra-Red Poles}

The first important check on the implementation is the verification of the
universal Infra-Red and Ultra-Violet poles in the dimensional regularisation
parameter $\eps$ \cite{Catani:1996jh,Catani:2000ef}. For the interfered amplitude in massless QCD these take a
rather simple form (in this case unrenormalised in the FDH scheme):
\begin{align}
  &Re(\mathcal{A}^{(0),\dagger}.\mathcal{A}^{(1),FDH,U}) =
  -\frac{1}{\eps^2}\left( N_c n_g + \frac{N_c^2-1}{2 N_c} n_q \right)| \mathcal{A}^{(0)}|^2 \nonumber\\&
  +\sum_{i=1}^{n-1} \sum_{j=i+1}^{n} \frac{1}{\eps} \log\left( \frac{\mu_R^2}{|s_{ij}|} \right)|\mathcal{A}^{(0)}_{ij}|^2
  -\frac{1}{\eps}\bigg( \beta_0
  +n_q \left( \frac{\beta_0}{2}-\frac{3}{2} \frac{N_c^2-1}{2N_c} \right)\bigg) | \mathcal{A}^{(0)}|^2 \nonumber\\&
  + \text{ finite terms,}
  \label{eq:poles}
\end{align}
where $n_q$ is the number of external quarks and $n_g$ is the number of
external gluons. $N_c$ is the number of colours and $n_f$ is the number of
light flavours and $\beta_0 = \tfrac{11N_c-2N_f}{3}$. The quantity
$|\mathcal{A}^{(0)}_{ij}|^2$ is the colour correlated Born amplitude defined by
\begin{align}
  |\mathcal{A}^{(0)}_{ij}|^2 = \mathcal{A}^{(0),\dagger} \cdot T_i \cdot T_j \cdot \mathcal{A}^{(0)}.
  \label{eq:born_cc}
\end{align}
where $T_g=if^{abc}$ and $T_q=T^a_{ij}, (T_\qb=-T^a_{ji})$. Scheme conversion to Conventional
Dimensional Regularisation (CDR) and renormalisation are all corrections
proportional to the Born amplitude and are defined as follows:
\begin{align}
  \frac{Re(\mathcal{A}^{(0),\dagger}.(\mathcal{A}^{(1),FDH,R}-\mathcal{A}^{(1),FDH,U}))}{|\mathcal{A}^{(0)}|^2} &= 
  -\frac{(n_g+n_q-2)}{2} \left( \frac{\beta_0}{\eps} - \frac{N_c}{3} \right) \\
  \frac{Re(\mathcal{A}^{(0),\dagger}.(\mathcal{A}^{(1),CDR,U}-\mathcal{A}^{(1),FDH,U}))}{|\mathcal{A}^{(0)}|^2} &= 
  -\frac{N_c}{3} - \frac{n_q}{4N_c} \left( \frac{N_c^2}{3}-1 \right) \\
  \frac{Re(\mathcal{A}^{(0),\dagger}.(\mathcal{A}^{(1),CDR,R}-\mathcal{A}^{(1),FDH,U}))}{|\mathcal{A}^{(0)}|^2} &= 
  -\frac{(n_g+n_q-2)\beta_0}{2\eps}
  -\frac{N_c}{3} - \frac{n_q}{4N_c} \left( \frac{N_c^2}{3}-1 \right)
  \label{eq:re}
\end{align}

\subsection{Accuracy}

We examine the accuracy of the amplitudes against the known pole structure and by using a re-scaling of the external momenta
to test the finite terms. $10^5$ phase-space points were generated using the RAMBO algorithm using rather weak kinematics cuts requiring
only that $p_i \cdot p_j > 10^{-4}s $ for a centre of mass energy of $s=7$ TeV.

\begin{figure}[h]
  \centering
  \begin{tabular}{cc}
    \psfrag{pp4j.0 virtual 2Re(A1.A0)}{\raisebox{-2mm}{\hspace{-5mm}\scriptsize $6g$}}
    \includegraphics[width=6.5cm]{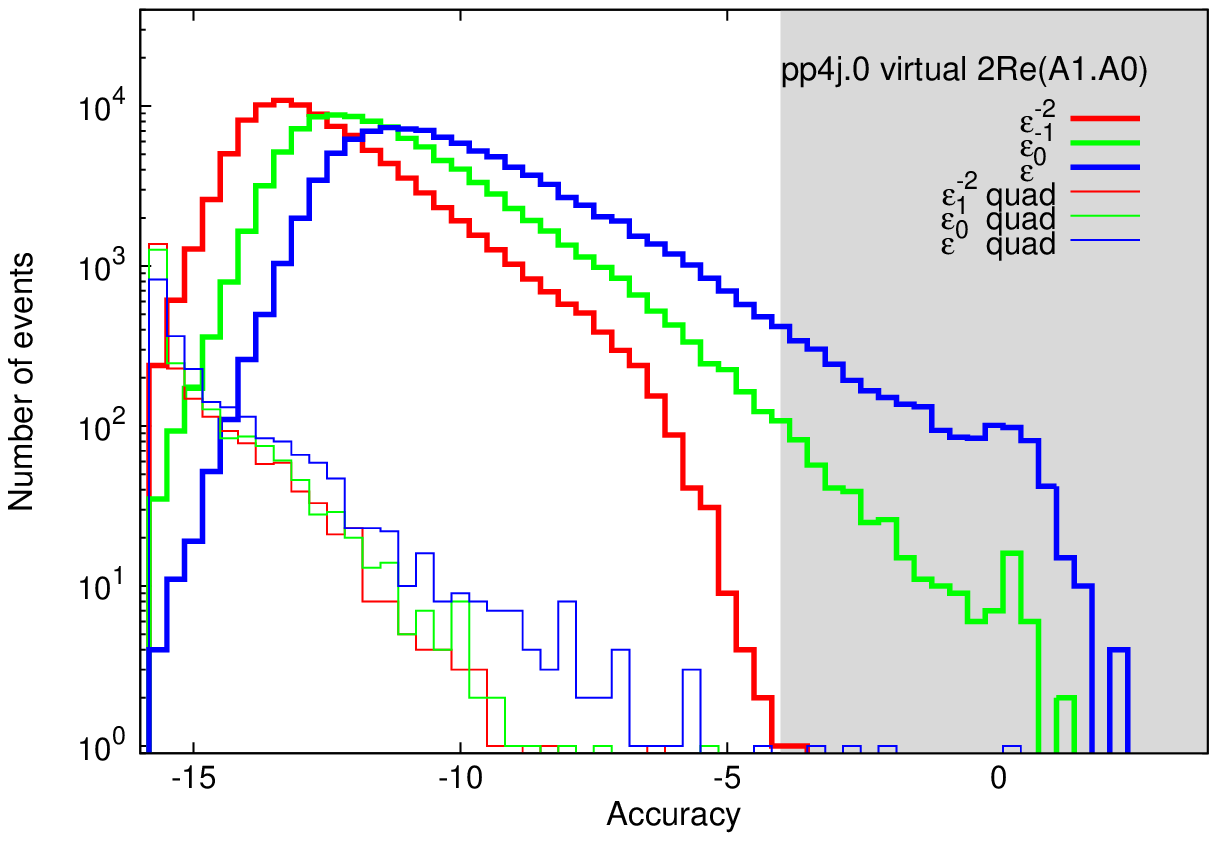} &
    \psfrag{pp4j.1 virtual 2Re(A1.A0)}{\raisebox{-2mm}{\hspace{-5mm}\scriptsize $\qb q + 4g$}}
    \includegraphics[width=6.5cm]{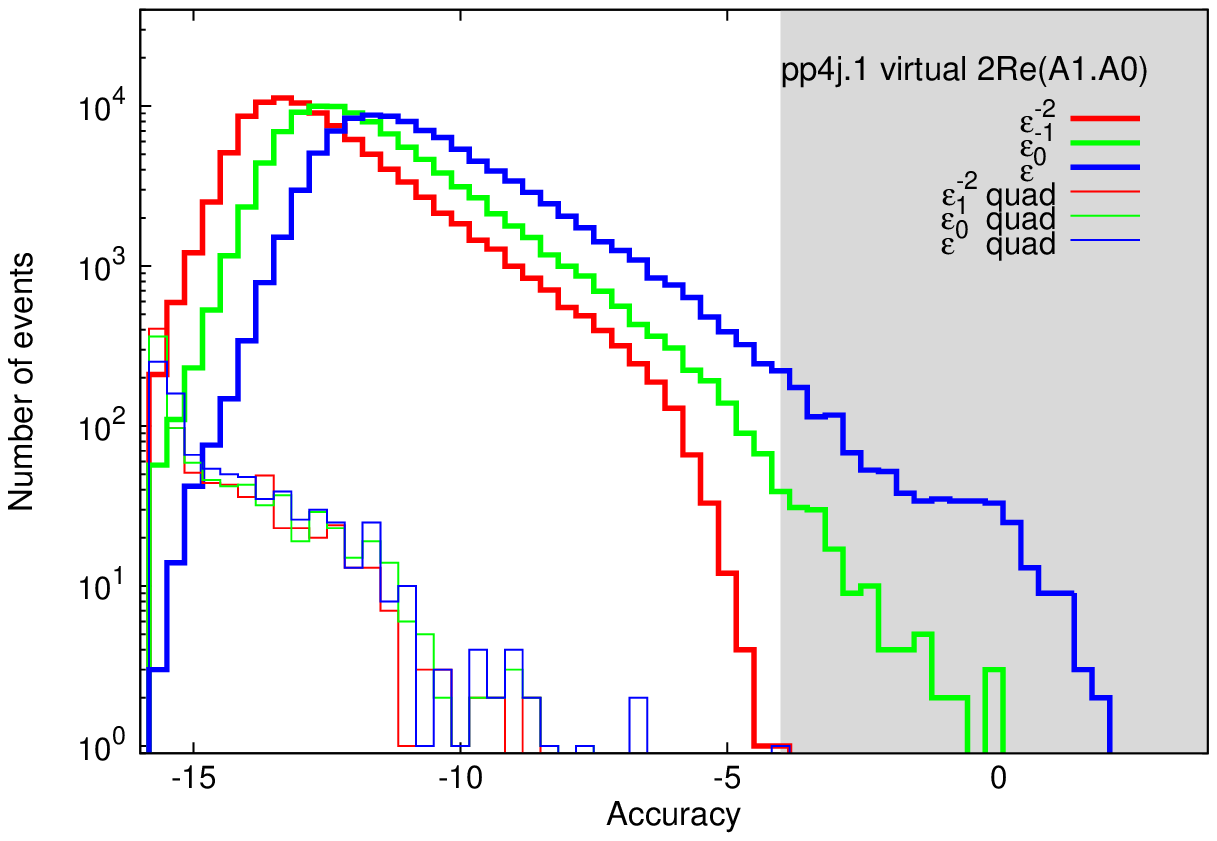} \\
    \psfrag{pp4j.2 virtual 2Re(A1.A0)}{\raisebox{-2mm}{\hspace{-5mm}\scriptsize $\db d \ub u + 2g$}}
    \includegraphics[width=6.5cm]{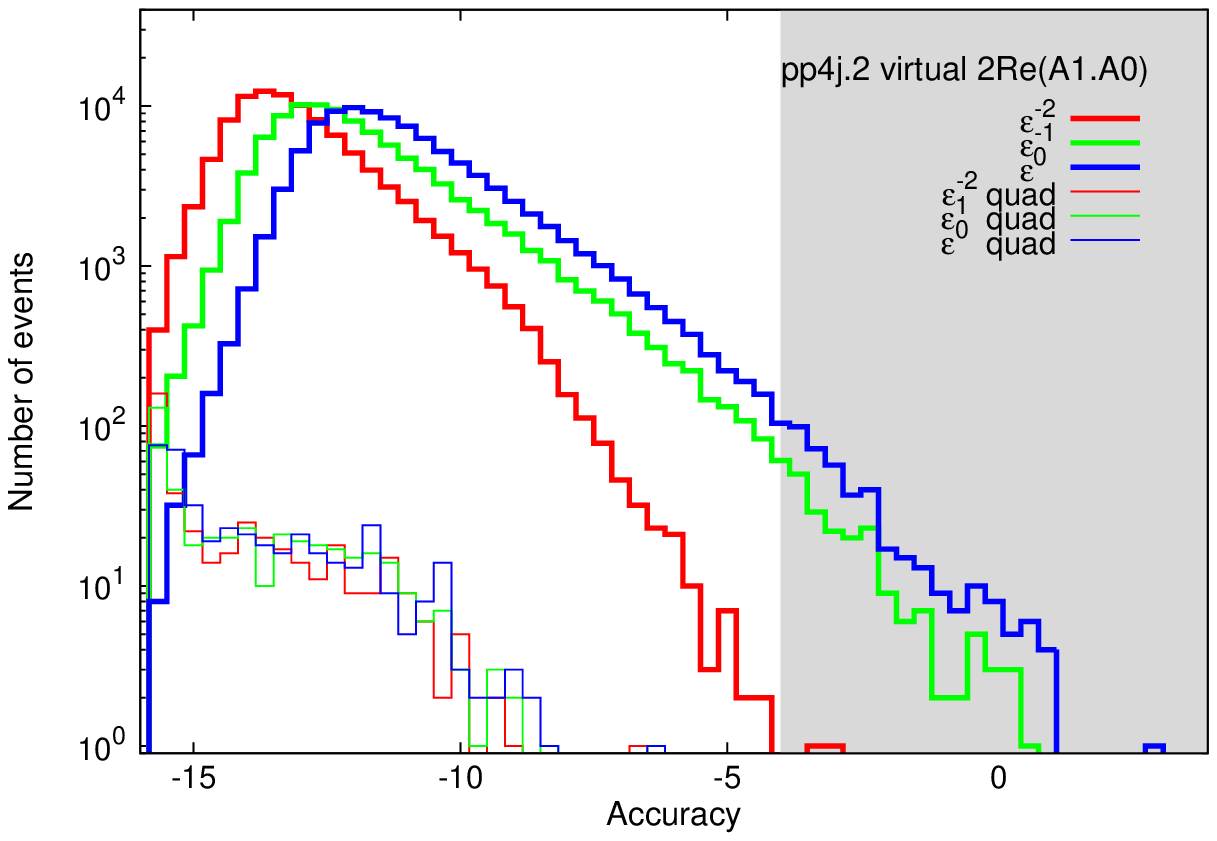} &
    \psfrag{pp4j.4 virtual 2Re(A1.A0)}{\raisebox{-2mm}{\hspace{-5mm}\scriptsize $\db d \ub u \sb s$}}
    \includegraphics[width=6.5cm]{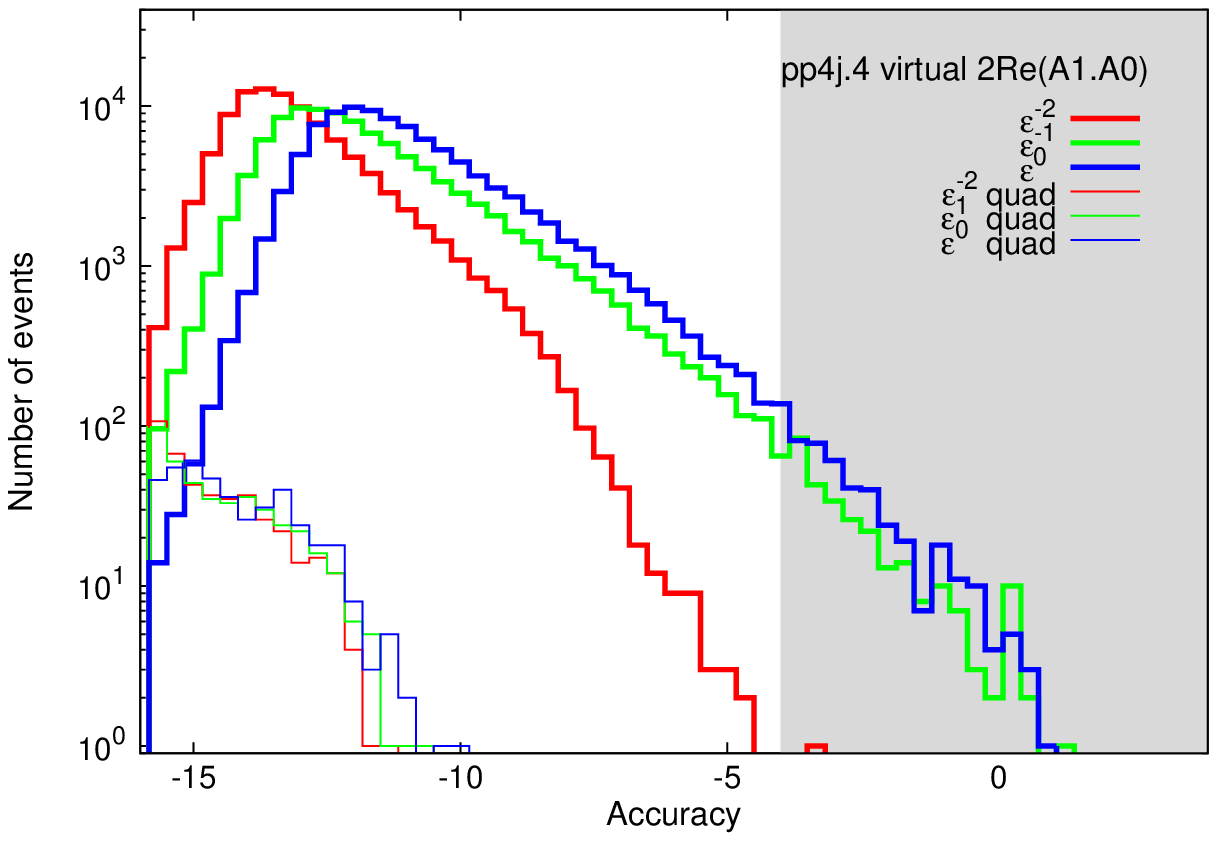}
  \end{tabular}
  \caption{Accuracy plots for the $2\rightarrow4$ processes contributing to the $4$-jet rate. Relative accuracy of $10^5$ points over a flat phase-space.
  Points appearing in the shaded region are re-evaluated in quadruple precision.}
  \label{fig:pp4j}
\end{figure}

The thinner lined histograms show the number of points re-evaluated in higher
precision (quadruple precision using the {\tt qd} package \cite{qd}).  In every
case, except for the six gluon amplitude, all re-evaluated points achieved the
required relative accuracy of $10^{-4}$. The number of points for re-evaluation
reaches a maximum of 2.2\% for six gluons and decreases rapidly with the
number of fermion pairs. In this most complicated channel 4 from 100,000 events
required octuple precision to reach the desired threshold, nevertheless they
would likely be excluded by common LHC cuts. Rough evaluation times for the
most complicated $2\rightarrow 4$ processes are of order $15-20s$ for full
helicity and colour and including re-scaling to compute an accuracy estimate
(see table \ref{tab:times}).

\begin{table}[t]
  \centering
  \begin{tabular}{|c|c|c|c|}
    \hline
    $n$ & 4 & 5 & 6 \\
    \hline
    $n$-gluons          & 0.03s & 0.5s  & 15.0s \\
    \hline
    {$q\qb+(n-2)$-gluons} & 0.03s & 0.75s & 17.0s \\
    \hline
  \end{tabular}
  \caption{Estimated evaluation times (s) for $n$-gluon and $q\qb+n$-gluon amplitudes using Intel Core i7 2.7GHz.}
  \label{tab:times}
\end{table}

\section{Conclusions}

We have described the implementation of full colour QCD virtual contributions
to NLO multi-jet production at hadron colliders. The primitive amplitude
decomposition and use of on-shell generalised unitarity shows to be a fast and
accurate method for the evaluation which scales well with the number of final
state particles. We are looking forward to some phenomenological applications 
in the near future.

\acknowledgments
We would like to thank Valery Yundin for many helpful discussions. We are also grateful to
Valentin Hirschi for providing some numerical cross checks obtained with MadLoop5 \cite{Hirschi:2011rb}.


\end{document}